\documentclass[twocolumn,aps,prd,amsmath,amssymb,nofootinbib]{revtex4}
\usepackage{amssymb,amstext,amsmath,amsthm}
\usepackage[dvips]{graphicx}
\usepackage{latexsym}
\usepackage{psfrag}
\usepackage{amsfonts}
\usepackage{bbm}

\newcommand{\Ref}[1]{(\ref{#1})}

\def\nn{\nonumber}

\newcommand{\eqa}{\begin{eqnarray}}
\newcommand{\neqa}{\end{eqnarray}}
\newcommand{\equ}{\begin{equation}}
\newcommand{\nequ}{\end{equation}}
\newcommand{\no}{\nonumber\\}

\def\la{\langle}
\def\ra{\rangle}
\newcommand{\bra}[1]{\la {#1}|}
\newcommand{\ket}[1]{|{#1}\ra}
\newcommand{\mean}[1]{\la{#1}\ra}

\newcommand{\p}{\partial}

\newcommand{\hh}{{\cal H}}

\def\d{\delta}
\def\f{\frac}

\usepackage{bbm}

\newcommand{\scr}{\rm\scriptscriptstyle}

\def\D{{\cal D}}

\let\eps=\epsilon
\def\vareps{\varepsilon}
\newcommand{\lp}{\ell_{\rm P}}

\newcommand{\Si}{\Sigma}
\newcommand{\si}{\sigma}
\newcommand{\Ga}{\Gamma}
\newcommand{\ga}{\gamma}

\begin{document}

\title{Background-free propagation in loop quantum gravity}

\author{Simone Speziale}
\email{sspeziale@perimeterinstitute.ca}
\affiliation{\small Perimeter Institute, 31 Caroline St, Waterloo, ON N2L 2Y5, Canada \linebreak
Centre de Physique Th\'eorique de Luminy, Case 907, F-13288 Marseille, EU}
\date{\small\today}
\begin{abstract}
I review the definition of $n$-point functions in loop quantum gravity, discussing
what has been done and what are the main open issues.
Particular attention is dedicated to gauge aspects and renormalization. 

\bigskip

\end{abstract}

\maketitle

\section{Introduction}
One of the tasks of quantum gravity is to provide a UV completion of the 
perturbative quantization in terms of gravitons, which unlike the other interactions in the Standard Model
turns out to be non-renormalizable.
It has been argued that the non-renormalizability is not intrinsic to general relativity itself,
but rather a problem of the perturbative approach. In particular, an aspect of the latter which is often criticized 
is the use of a fixed background 
that is needed to have a quadratic leading order in the action and thus to
be able to perform the perturbative quantization.
From this point of view, it is suggestive the example of general relativity in 2+1 dimensions.
This theory (in the first order triad formalism) 
has a quadratic leading order without need of introducing a non-zero background; the
perturbative expansion can be constructed around the zero classical value of the field and it is 
renormalizable\footnote{The intuitive reason for the renormalizability is that in 2+1
$R_{\mu\nu}=0$ implies $R_{\mu\nu\rho\sigma}=0$, thus all countertems vanish on-shell.}
 \cite{Witten}.

Loop quantum gravity \cite{books} pursues this line of thoughts, making of background-independence the guiding principle
for the quantization of general relativity.
The result is a mathematically sound theory where the Planck length emerges as a dynamical
scale at which spacetime becomes granular and discrete. Despite this appealing microscopic picture,
the low-energy interpretation of the theory is less clear: in a snapshot, 
the main open issue is to derive the low-energy
approximation starting from the basic non-perturbative formalism.

The last three years have seen important progress in this direction, and this is the basis of this review.
A technique has been introduced and developed to study $n$-point functions within loop quantum gravity,
thanks especially to the work of Rovelli \cite{Conrady, Modesto, Rovelli, Bianchi}.
This technique offers an explicit framework for extracting physics
-- in particular, it allows us to define a perturbative expansion in $\lp$ and thus test
the low-energy interpretation of the theory.
If the correct semiclassical limit arises, 
this technique can explore the way loop gravity UV-completes the linearized quantum theory of gravitons.

The aim of the present non-technical review is to introduce the reader to this technique and give a broad presentation
of some of the results obtained so far in its application to loop quantum gravity.
In the next Section, I will discuss the general set up to
describe propagators and $n$-point functions in a fashion extendable to background-independent theories.
In Section \ref{SecSF}, I discuss the application of this set up to the spin foam formalism for loop quantum gravity.
The material of these two initial Sections is mainly based on Rovelli's intuition, and gives my perspective on what appeared in \cite{Conrady, Modesto, Rovelli, Bianchi}. New is the attention given to gauge aspects and
renormalization.
In Sections \ref{Sec3d} and \ref{Sec4d}, I report on some of the explicit results that have been obtained in
specific spin foam models for three and four dimensional quantum gravity. This part is based on work
of my collaborators and myself, appeared in \cite{Rovelli, Bianchi, Livine, Chris, 3d, 3d1, 3d2, bs}.
The final Section \ref{SecConcl} contains my conclusions, an overview of what has been achieved so far
and what I believe are the most relevant steps to take next.

\section{General boundary correlations}
Consider perturbative quantum gravity around Minkowski,
$g_{\mu\nu} = \eta_{\mu\nu} + h_{\mu\nu}$. It is useful to recall that in the linearized theory
the spatial diffeo constraints of the full theory are still present, 
whereas the Hamiltonian constraint splits into a constraint (morally the tracelessness condition)
plus a true Hamiltonian. 
The standard definition of the graviton propagator, or 2-point function, 
involves a functional integration over the whole spacetime,
\equ\label{1}
\bra{0} h_{\mu\nu}(x) h_{\rho\sigma}(y) \ket{0} = \int \D h_{\alpha\beta} \, h_{\mu\nu}(x) \, h_{\rho\sigma}(y) \, e^{i S[h]}
\nequ
Consider now two hyperplanes $\Si_i$, $i=1,2$ located respectively at time $t=0$ and $t=T$, 
and assign field values $h_{ab}{}^i$ on them (here $i=1,2$ and $a=1,2,3$).
The propagation kernel
\equ\label{K}
K[h_{ab}{}^1, h_{ab}{}^2, T] = \int_{h_{ab}{}^2 \atop h_{ab}{}^1} \D h_{\alpha\beta} \, e^{i S[h]}
\nequ
can be evaluated perturbatively in the temporal gauge $h_{0\mu} = 0$.
This non-covariant gauge allows us to bridge between spacetime path integrals
and the canonical formalism: the propagation kernel satisfies
\equ\label{K1}
K[h_{ab}{}^1, h_{ab}{}^2, T] = \sum_n e^{-i E_n T} \, \overline{\Psi_n[h_{ab}{}^1]} \, \Psi_n[h_{ab}{}^2],
\nequ
where the $\Psi_n$ are a complete set of physical states, 
namely they satisfy spatial diffeos and Hamiltonian constraints,
and are eigenstates of the true Hamiltonian. Although $h_{0\mu} = 0$ is a priori only a partial gauge fixing,
the resulting expression is fully diffeomorphism invariant. See \cite{Mattei} for details.

The correlator \Ref{1} can be written in terms of the kernel \Ref{K} splitting the functional integration $\D h$
into five regions, characterized by the following $t$ intervals: 
$(-\infty,0)$, $t=0$, $(0, T)$, $t=T$ and $(T, \infty)$. The $(0, T)$
integration can be directly identified with the definition \Ref{K} of the kernel. 
As for the $(-\infty,0)$ and $(T, \infty)$ integrations, notice that sending one time extremum
to infinity in \Ref{K} amounts to projecting onto the minimal energy state $\Psi_0[h]$, as can be seen
from (the analytic continuation of) \Ref{K1}. 
Consequently, the 2-point function can be written
\eqa\label{W}
\bra{0}h_{ab}(x) h_{cd}(y)\ket{0} =&& \\\nn &&\hspace{-3.2cm}
= \int \D h^1 \, \D h^2 \, \Psi_0[h^1] \, \Psi_0[h^2] \, K[h^1, h^2, T] \, h_{ab}(x) \, h_{cd}(y).
\neqa
Like \Ref{1}, this expression depends on the background $\eta$ with respect to which 
the otherwise meaningless points $x$ and $y$ are identified.\footnote{See \cite{Bianchi} for a discussion
of the difficulties of defining $n$-point functions in a background-independent context.}
It can be gauge-fixed\footnote{Due to the field insertions,
the additional gauge-fixing of time-independent spatial
diffeos is now needed. This can be achieved for instance using the Coulomb gauge $\p_a h^a{}_b = 0$.} 
and evaluated perturbatively.

The vacuum state $\Psi_0$ thus introduced is a key object of the theory. 
Little is known about the properties of the true vacuum state of full non-perturbative quantum gravity.
Yet in the case at hand where we are only interested in recovering the theory of gravitons,
we can limit ourselves to zero cosmological constant and asymptotically flat boundary conditions.
Under these circumstances, Minkowski is the minimal energy state.
In the linearized theory, a true Hamiltonian is available and 
can be used to identify and evaluate the minimal energy state,
which turns out to be a Gaussian peaked on Minkowski space. 
Schematically, $\Psi_0[g] \sim \exp\{-\int (g_{ab}-\d_{ab})^2\}$ \cite{Kuchar, Mattei}.
Furthermore, the positive-action theorem \cite{stability} suggests that it is unlikely that instabilities arise 
even in the full theory, thus $\Psi_0$ in \Ref{W} can be identified with the Minkowski state 
non-perturbatively.\footnote{The situation of course changes if one has
non-trivial boundary conditions, or if non-zero temperature is allowed. These interesting situations
are beyond the scope of this review.}

\subsection{General Boundary}
In the (perturbative) evaluation of functional integrals like \Ref{K} or \Ref{W}, spatial asymptotic
conditions have to be specified, and vanishing $h_{ab}$ are the ones used in the graviton theory.
Hence in \Ref{K} we can imagine that the full set of boundary data is $h_{ab}{}^1$ and $h_{ab}{}^2$ on the two hyperplanes,
plus $h_{ab}=0$ at spatial infinity for any time in between. This set of boundary data lives on
a 3d surface, obtained by the union of $\Si^1$ and $\Si^2$ with the timelike
boundary at spatial infinite. Formally, we can imagine to make this whole 3d surface, denote it $\Si$, \emph{finite}
and closed.
Let me first consider the case of Riemannian signature.
Coordinate $\Si$ with $x^a$, $a=1,2,3$, denoting $x_4$ the ``radial'' coordinate in the bulk, and simply
$h_{ab}$ the classical field assigned to the surface. 
Define the kernel 
\equ\label{Kgb}
K[h_{ab}] = \int_{h_{ab}}\D h_{\mu\nu}\, e^{i S[h]}
\nequ
as the functional integral inside $\Si$ with boundary value $h_{ab}$ on $\Si$ fixed,
in the radial gauge $h_{4\mu}=0$.\footnote{This radial gauge is 
the analog of the temporal gauge for \Ref{K}, and it is again only a partial gauge choice. 
The possibility that the kernel is again fully gauge fixed was discussed in \cite{Magliaro}.}
In Lorentzian signature, the value of the gravitational field 
along the timelike part of $\Si$ would play the role of $T$ in \Ref{K}.
Formally this expression is (up to the $i$ in the exponent) the familiar Hartle-Hawking state \cite{HH}.
Like the kernel in the standard form \Ref{K}, this functional of $h_{ab}$ formally satisfies
the Hamiltonian constraint (or Wheeler-DeWitt equation), as a consequence of the integration over the diffeomorphisms
that imposes the spatial and Hamiltonian constraints \cite{Mattei}.

It is natural to expect \Ref{Kgb} to be peaked on the classical solution associated to the boundary 
data $h_{ab}$, although admittedly we have little knowledge about the classical properties (existence, uniqueness) of this kind of general boundary problem (see e.g. \cite{Pfeiffer}).
In general, the kernel gives a probability amplitude for the metric $h_{ab}$ on $\Si$ obtained
integrating over all the interactions inside $\Si$. It can be used to compute $n$-point functions:
using this finite $\Si$ we can formally rewrite \Ref{W} as 
\equ\label{W1}
\bra{0}h_{ab}(x) h_{cd}(y)\ket{0} = \int \D h \, \Psi_0[h] \, K[h] \, h_{ab}(x) \, h_{cd}(y)
\nequ
where $\Psi_0[h]$ is now the vacuum state on the (connected) closed surface $\Si$, defined as the functional
integration outside of $\Si$. I will talk extensively
about this state in the next Section. For the moment, 
let me stress that this equation is just a formal manipulation of the usual definition \Ref{1} of
the propagator. Its (gauge-fixed) perturbative expansion yields the same non-renormalizable theory,
albeit with the increased difficulty of evaluating the various quantities of interest on a closed $\Si$.
Notice also that for Lorentzian signature the metric $h_{ab}$ on $\Si$ will have coordinate singularities.

\subsection{Semiclassical boundary states}
Let us focus on the general boundary case where $\Si$,
instead of being simply the union of $\Si^1$ and $\Si^2$ with the timelike
boundary at spatial infinite, is a generic 3d surface. 
In \Ref{W1} the kernel and vacuum state now depend on $\Si$ through the metric induced on it
by the  Minkowski background. This induced metric, denote it $q_{ab}$, is not trivial anymore, and 
it can be as arbitrary as the deformations of $\Si$ itself. 
There are now many possible vacuum states $\Psi_0[h, \Si]$, depending on the location
of $\Si$ with respect to the Minkowski background or, equivalently, on the classical 
metric $q$.
Each vacuum state $\Psi_0[h, \Si]$ defines a state $\Psi_q[g]$ for the full metric 
$g_{ab} = q_{ab} + h_{ab}$.
Just as for the case of trivial $\Si$, it is reasonable to assume that $\Psi_q[g]$ is 
peaked on $q$.\footnote{This argument can be visualized explicitly for the linearized theory, 
where we expect schematically $\Psi_q[g] \simeq \exp -\int(g_{ab}-q_{ab})^2$,
generalizing the standard case where $\Si$ is an hyperplane and 
$\Psi_{q=\d}[g] \simeq \exp -\int(g_{ab}-\d_{ab})^2$.
I need to add that while this expectation is perfectly reasonable, the explicit calculation,
following for instance the Green's function method used in \cite{Mattei} for the
hyperplane case, is technically challenging, and explicit solutions for generic 
$\Si$ are not known, to the best of my knowledge, even for free theories.}
The definition
\equ\label{P}
\Psi_q[g] \equiv \Psi_0[h, \Si], \qquad g_{ab} = q_{ab}+h_{ab},
\nequ
allows a key shift of perspective: $\Psi_0[h, \Si]$ is a functional integral over the whole of
spacetime outside $\Si$ (with vanishing asymptotic boundary conditions), which results 
in a state where the full metric $g$ is peaked on $q$. 
We can thus simply think of $\Psi_q[g]$ as a semiclassical state
peaked on the metric $q$, independently of the Minkowski background. 
In a background-dependent context, the dependence on $q$ simply reflects the dependence on the location
of $\Si$ with respect to Minkowski. In a background-independent picture, there is no location in spacetime:
the dependence on the boundary geometry is not in the location of $\Si$, but on the value $q$ of the
gravitational field itself.
In the words used above, the multiplicity of possible
locations with respect to the Minkowski background in the background-dependent viewpoint translates into
a multiplicity of possible semiclassical boundary states in a background-independent viewpoint.

The metric $q$ gives meaning to the spacetime points in the evaluation of the $n$-point functions.
The latter should be thought of as expectation values on the given semiclassical state $\Psi_q$,
i.e. $W^q_{abcd}(x,y) \equiv \bra{\Psi_q} h_{ab}(x) h_{cd}(y) \ket{\Psi_q}$.

Given the key role of the boundary state, let me spend a few more words about it.
First of all, it has to be a dynamical state: by its very definition, it is, like the kernel,
a solution of the Hamiltonian constraint.
I said that it is natural to assume that $\Psi_q[g]$ is peaked on the metric $q$. As Rovelli pointed out in
\cite{Rovelli}, a good semiclassical state should be peaked on both intrinsic and extrinsic geometry, just as a
semiclassical state in quantum mechanics is peaked on both position and momentum.
This is indeed the case from its definition \Ref{P}: as soon as $\Si$ is not trivial,
there is a boundary term in the action which gives a non-vanishing contribution to the 
evaluation of the functional integral. This boundary term is precisely
the extrinsic curvature of $q$ embedded in Minkowski \cite{Gibbons}. 

What I have said so far is purely formal. In particular, there is no well-defined notion of Wick rotation,
the integration measures are ambiguous and the UV divergences of the perturbative expansion can not be renormalized. I now turn to discuss how these ideas can be applied to a well-defined
background-independent theory, loop quantum gravity.

\section{Spin foam correlations}\label{SecSF}

Adapting \Ref{W1} to loop gravity requires taking into account the fundamentally discrete 
nature of the theory. In canonical loop quantum gravity, the metric is an operator, and the kinematical 
(i.e. prior to the imposition of the Hamiltonian and spatial diffeo constraints) Hilbert space $\hh_0$ is spanned 
by spin networks $\ket{s}$,
where in $s=(\ga, j_l, i_n)$ $\ga$ is a graph, $j_l$ a set of half-integers associated to the links of the graph, and 
$i_n$ an additional set of half-integers associated to its nodes.
These two sets of quantum numbers come from the group theory of SU(2), whose relevance to gravity
has been widely discussed in the literature: the $j$'s are the spins labelling the
irreducible representations (irreps) of SU(2), and the $i$'s are the interwiners projecting tensor
products of irreps into the gauge-invariant (singlet) ones.
The spin network basis diagonalizes kinematical geometric operators such as areas and volumes of 
coordinated regions of space. This means that at the kinematical level, the possible outcomes
of the gravitational field are labelled by these states.

To study the dynamics canonically, one imposes the constraints  \`a la Dirac
as restrictions on the kinematical states: the states in $\hh_0$ annihilated by the constraints define
the physical Hilbert space $\hh_{\scr Phys}$.
Thanks to the efforts of Ashtekar, Lewandowski and Thiemann among others, this program has achieved important results. These include the full imposition of the
spatial diffeo constraint, resulting in the space of so-called abstract spin networks, and a UV-finite
rigorous definition of the quantum Hamiltonian constraint.
However the latter is not free of ambiguities, this problem being also reflected in a non completely satisfactory
control over the quantized version of Dirac's constraint algebra.
Furthermore, although an infinite number of solutions of all constraints was found by Rovelli and Smolin \cite{RS},
a complete characterization of physical states seems still out of reach.
Due to this type of difficulties, a number of researchers have moved their attentions to 
an alternative approach to the dynamics, known as the spin foam formalism, through 
the construction of spin foam models conjectured to implement the Hamiltonian constraint in a covariant way.

This conjecture is rather solid in 2+1 spacetime dimensions, 
and recent important progress has been also made in 3+1. I will come back to these points below.
For the moment, it suffices to say that the spin foam formalism is a covariant ``sum-over-histories'' version of loop gravity: 
Given initial and final spin networks $s$ and $s'$ representing the boundary gravitational field, 
a spin foam history $\si$ is a 2-complex $\Ga$ interpolating between the
two graphs, with faces labelled by spins and edges by intertwiners;
a (model-dependent) weight ${\cal A}_\si[s, s']$ is associated to each such history,
and the quantum amplitude $K[s, s']$ is obtained summing over all the spin foams compatible with the boundary data,
\equ\label{Kss}
K[s,s'] = \sum_{\si|\p\si = s\cup s'} {\cal A}_\si[s,s'].
\nequ
The quantum amplitude encodes the full spin foam dynamics, and provides a non-perturbative and background-independent definition of the kernel \Ref{K}. 
See \cite{books} for a more complete introduction and references.
The integration over 4-geometries is realized in \Ref{Kss} by the summation over internal spin foams.
The latter means a summation over all the possible 2-complexes $\Ga$ times a summation
over all the possible labellings in terms of spins and intertwiners. For the case when
$\Ga$ is dual to a simplicial manifold, the second summation corresponds to a summation 
over all the possible (discrete) metrics associated with it.

The actual summation is tentatively implemented through a generalization of matrix models that goes under
the name of group field theory \cite{DePietri}. The latter generates naturally all possible cellular two-complexes,
which are not necessarily dual to triangulations. This makes the summation extremely
rich, and extremely hard to control \cite{DePietri1}.
Each 2-complex is weighted by $\lambda^V$, where $\lambda$ is the coupling constant
of the theory and $V$ the total number of vertices in the 2-complex.
Hence the amplitude ${\cal A}_\si$ appearing in \Ref{Kss} depends on $\lambda$.
This coupling constant controls the summation over the 2-complexes.

For applications to \Ref{W1}, we consider the case when the boundary is connected and the kernel is a 
function of the only boundary spin network $s$,
\equ\label{Ksf}
K[s] = \sum_{\si|\p\si = s} {\cal A}_\si[s].
\nequ
In a consistent model the kernel projects on the physical Hilbert space of loop gravity,
as one would demand of \Ref{Kgb} in the continuum. 

If loop gravity describes the correct physics, 
a formula like \Ref{W1} is expected to give the right semiclassical limit 
with a sensible and divergence-free UV completion. 
But how can we translate \Ref{W1} to loop gravity? First of all,
because the Hilbert space of possible values of the gravitational field is labeled by spin networks,
the integration over 3-metrics is replaced by a summation over spin networks.
As discussed above, \Ref{W1} can be used in a background-independent theory if
the state $\Psi_q$ is interpreted as a (observed-dependent) semiclassical state peaked around the geometry $q$, 
and the points $x$ and $y$ are identified with respect to $q$.
In the following, I will make the important assumption that loop gravity admits
a semiclassical state with the properties discussed above, namely to be a function of a 3-metric 
$g$ peaked on a given intrinsic and extrinsic geometry, which we denote shortly with $q$.
This state is not required to be the vacuum state of the full theory,
but it has to be physical, i.e. a solution of the Hamiltonian constraint.
Under this assumption, and choosing a specific spin foam model for $K[s]$,
the  correlator \Ref{W1} can be realized 
identifying (i) the boundary Hilbert space on $\Si$ with the Hilbert space of abstract spin networks; (ii) the field insertions $h_{ab}$ with corresponding
expectation values of canonical operators in loop quantum gravity, 
${\mathbbm h}_{ab}\equiv \bra{s}{\hat h}_{ab}\ket{s}$; and (iii) the boundary state with a suitable spin network functional $\Psi_q[s]$ peaked on the classical geometry $q$. 
Namely,
\equ\label{Wsf}
W^q_{abcd}(x_1, x_2) = \sum_{s} \, {\mathbbm h}_{ab}(x_1) \, {\mathbbm h}_{cd}(x_2) \, \Psi_q[s] \, K[s]. 
\nequ
See e.g. \cite{Rovelli,Bianchi} for a more exhaustive description.

In the continuum theory, the graviton
propagator corresponds to the correlation between the excitation of a quantum of the gravitational field at a
point $y$, given a quantum at the point $x$. In the loop discrete setting, it corresponds to the
correlator between excitations of quanta of space. The operators associated to the metric are of two
different types, which can be interpreted as areas and dihedral angles. They are associated to,
respectively, diagonal and off-diagonal components of the metric tensor.

Can the spin foam definition \Ref{Wsf} of the graviton propagator \Ref{W} 
yield a unitary and renormalizable, if not finite, quantum theory?
The heuristic reason to expect UV finiteness is that, similarly to string theory, 
interactions are not anymore point-like. 
However, while in the latter this is due to the extended (stringy, brany) nature
of things, in loop gravity it occurs because spacetime itself has a granular structure at the Planck scale:
the geometric operators have discrete spectra with minimal eigenvalues proportional to $\lp$.
In the simplest cases, these ``quanta of space'' can be thought of as Regge cells with only
discrete values of the geometry allowed. 
Yet notice that the fundamental discreteness of space in loop gravity is very different from having a lattice:
it is not an assigned fixed property of space, but a consequence of the spectrum of
the geometric operators, manifest in the spin network basis. 
For instance, a striking difference with lattice theories is that 
Lorentz symmetry is compatible with loop gravity \cite{noiLorentz}.

If this heuristic picture is implemented dynamically by a spin foam model
we expect the calculation of correlators such as \Ref{Wsf} and scattering amplitudes to be UV finite.
In this perspective, it is remarkable that finite 4d Lorentzian models are explicitly known \cite{Crane}.
Notice that in this scenario the UV completion of general relativity does not require any new physics:
it is rather a property of the theory itself. In this sense, this approach is at the same time
conservative and very ambitious.
 
\subsection{Perturbative expansion}
The definition \Ref{Wsf} is completely background-independent; the full gravitational field
is quantized. Yet a classical metric $q$ enters crucially the expression, through
an observer-dependent semiclassical boundary state. This allows us to 
introduce a perturbative expansion in powers of $\lp$ around $q$.
In turn, the perturbative expansion allows us to study and test the semiclassical regime of the chosen spin foam model:
in the continuum theory the leading order of the graviton propagator 
encodes Newton's law and thus the correct classical behaviour of the theory. 

How do we define the perturbative expansion of \Ref{Wsf}? First of all, 
in loop gravity the spectrum of geometric operators scales like $j \lp$.
This means that the $\lp$ expansion can be studied taking the large spin limit at fixed $j \lp$.
Recall that in the SU(2)-based 
quantum theory of angular momentum the large spin limit corresponds to the semiclassical limit,
i.e. the coupling of vectors in flat 3d space. It should then not come as a surprise
that classical geometry emerges in the large spin limit of spin foams. 
The relevance of SU(2) and of its semiclassical limit is clear
for three dimensional euclidean gravity, where (a) SU(2) is indeed the gauge group of
general relativity, and (b) gravity is about flat space. 
But why should it be relevant to 4d as well?
Indeed, the basis of loop gravity was the discovery that
(a) still applies, thanks to the variables introduced by Ashtekar \cite{Ashtekar}:
an SU(2) connection and its conjugate field which represents the metric.
A key feature of Ashtekar's connection is to include the extrinsic curvature.
The latter is responsible at the quantum level for non-trivial deficit angles between flat chunks
of space. This is where part (b) becomes non-trivial \cite{Immirzi}. 
How curved geometry is dynamically encoded is still unclear, and it is part of the problem
of understanding the semiclassical limit of 4d spin foam models.
I will come back to this in Section \ref{Sec4d} below.

Technically, the $\lp$ perturbative evaluation can be set up using 
the SU(2) harmonic analysis to re-express the sums over spins as integrals over the group \cite{Livine}.
In doing so the spins end up in the argument of exponentials, and
the large spin expansion amounts to the asymptotic evaluation of the integrals using saddle point techniques.

How about the dependence on $\lambda$, the coupling constant of the group field theory? 
By this construction, the summation over spin foams entering the correlators is a power series in 
$\lambda$. A priori, there is no reason to expect
$\lambda$ to be related to $\lp$, so this furnishes \Ref{Wsf} with a second and genuinely independent
perturbative expansion. What is the meaning of this expansion? 
Naively one could expect higher orders in $\lambda$ to include shorter scale corrections. However 
a fixed triangulation can carry both a very large or a very small geometry:
as Rovelli remarked (e.g. \cite{EPR}), the number of $n$-simplices in a simplicial manifold is not a IR nor
a UV cut-off, but rather a cut-off on the ratio 
between the overall size of the spacetime region considered and the smallest wavelength allowed.
This suggests that given a process with typical scale $\mu$, there is an optimal ratio among
the number of $n$-simplices and their average size, and triangulations away from this ratio are subdominant.
I would like to borrow a simple
and elegant scenario from renormalization theory, where all these subdominant contributions can be effectively
absorbed in the rescaling $\lambda(\mu)$ of the coupling constant. 
For instance, denoting $N$ the number of 4-simplices in the bulk,
$\lambda(\mu)$ should scale in such a way that the expectation value
$N\big(\lambda(\mu)\big)$ well describes the physics at the scale $\mu$.
Such a scenario offers what I fear is one of the very few ways that we can hope
to be able to tame the most difficult combinatorial problem
of summing over all the 2-complexes. I will come back to this point below.

Alternatively, a point of view especially advocated by Freidel \cite{Freidel} is that only the
simplest triangulation should be included in the bulk. This is equivalent to saying that all the other configurations
are just elements of the gauge orbits.

Finally, a very different scenario has been sponsored by Markopoulou and Oriti among 
others (see e.g. \cite{Konopka, Oriti}), where 
the semiclassical properties, diffeo invariance and field equations, are absent in the
fundamental theory and only emerging statistically.

Here I consider only the first case. The state of the art is that we have a fair understanding of the
fixed triangulation dynamics, whereas the sum over triangulations bears a number of open questions.

\subsection{Diffeomorphism invariance and gauge fixing}

Before moving to explicit models, it is important to 
discuss how the loop approach defines the quantum measure.
The 3d diffeomorphism invariant measure appearing in \Ref{W1} is realized as a sum over the spin network states,
\equ\label{D3}
\int \D g_{ab} \mapsto \sum_s = \sum_{\ga} \sum_{j_l, i_n}.
\nequ
The first is a summation over all the graphs, and the second over all the possible metrics
associated with each graph. 
Similarly, also the covariant 4d measure is defined by a summation over all spin foams,
\equ\label{D4}
\int \D g_{\mu\nu} \mapsto \sum_{\si} = \sum_{\Ga} \sum_{j_f, i_e}.
\nequ
Notice that a priori this type of summations do carry gauge redundancy, in the form of different
configurations with the same value of the summand. My intuition is that in general a spin foam
model will carry gauge degrees of freedom, and only once these are fixed the model can yield finite answers.
Of course, this intuition does not prevent the possibility, envisaged by some authors in the community,
that a clever enough spin foam model could capture directly only the physical degrees of freedom.
In any case, the latter property would clearly belong to a very specific type of model. For the
general case I believe it is instructive to discuss the gauge structure of the spin foam summations. 

The gauge structure can be understood by analogy with \Ref{W1}.
In \Ref{W1}, the radial gauge is used to fix the kernel and boundary state, and an additional
gauge-fixing of $x_4$-independent boundary diffeos is needed because of the gauge-dependent field insertions.
It is natural to expect a similar structure in \Ref{Wsf}. Namely, the summation over spin foams
\Ref{D4} to be gauge-fixed with a radial gauge and consistent with the boundary state.
Then, also the sum over spin networks \Ref{D3} should be gauge-fixed if one is evaluating observables
depending on boundary diffeos.

Where is the gauge redundancy? Both boundary \Ref{D3} and bulk \Ref{D4} summations have a double nature:
first, as a sum over the possible metrics (represented by spins and interwiners)
associated to a given graph. Second, as a sum over all possible graphs.
Naively, gauge degrees of freedom will be present in both types of summations. 

Consider first a fixed graph dual to a triangulation.
This choice allows us to use the experience gained from lattice models of gravity to investigate the
gauge redundancy of the sums over the spin labels.
Perturbative lattice Regge calculus around flat spacetime teaches us that there are three types of variables
in the discrete path integral: physical degrees of freedom, gauge variables, and spurious variables
which are just lattice artifacts, and decouple and vanish in the continuum limit.
The identification of the spurious variables is transparent only when dealing with
perturbations around flat spacetime on a regular lattice, but is in general 
a potentially ambiguous procedure.
Once identified, these variables can be safely fixed to zero. The remaining gauge freedom
lies in the possibility of arbitrarily deforming the lattice by moving a vertex around in the flat
background. Fixing this gauge can be done by a direct discretization of the continuum procedure. 
See \cite{Rocek, Hamber1, Dittrich} for details.\footnote{
In 3d quantum gravity without matter, where curved configurations are not even allowed off-shell, 
this gauge freedom extends to the non-perturbative level. 
}

The situation changes significantly for curved spacetime. While in the continuum case we have
an extended understanding of diffeomorphism invariance, I am not aware of any 
rigorous classification of physically equivalent discrete curved manifolds, which
is needed for a general approach to gauge-fixing.
A priori one is tempted to imagine that, unlike in the flat case, 
moving around a vertex in a curved lattice genuinely changes
the metric and does not correspond to a discrete diffeomorphism anymore.
Nevertheless, gauge invariance can still be expected. As a simple example, consider
the special case where a vertex is sitting inside a flat Regge cell (or $n$-simplex): then
any movement of the vertex which keeps it inside the $n$-simplex corresponds to a gauge transformation.
Because this situation can always be obtained with a suitable refinement of any initial triangulation,
it is tempting to conclude that whenever there is a redundant refinement  
gauge degrees of freedom are associated with perturbations around it.
This argument is far from giving us a real understanding of gauge-fixing on a curved lattice, but
it shows that the issue should not be neglected.\footnote{It also
shows how the dimensionality of the gauge orbits of the diffeomorphism group is not
constant. This peculiar aspect is typical of the nature of the diffeomorphism group,
and it is also present in the continuum case. For instance when
the space manifold is compact the dimensionality of the orbits changes
with the existence of Killing vectors (see e.g. \cite{Moncrief} for a review).}

Given the limited control that we have in identifying gauge-equivalent configurations 
on a fixed triangulation, the reader might anticipate that
the situation get worse when we look at the summation over 2-complexes. 
The only case which I think I understand concerns topological field theories
(i.e. theories without local degrees of freedom), such as 3d general relativity, 
and even in this case I will only consider sums over all the triangulations,
rather than over all the 2-complexes.
In this restricted setting there is a powerful method to identify gauge-equivalent 
configurations: any two triangulations related by a sequence of Pachner moves \cite{Pachner}
have the same topology. These moves can thus be used to explicitly construct gauge-equivalent
contributions to the evaluation of correlators, and I will come back to this point in the
next Section, where I describe the 3d case more in details. 

In any case, because a priori I can not rule out the possibility that also a generic 2-complex
can carry gauge degrees of freedom, I think that in spite of the help coming from
lattice gravity models, the issue of gauge fixing can be completely understood only at the
level of the group field theory that generates the sum over 2-complexes.

In the rest of the paper, I will discuss explicit spin foam models and review the results appeared in the literature
for their correlators. 
All the literature so far deals only with Euclidean signature. I will comment to the Lorentzian
case in the conclusions.

\section{Three dimensional case}\label{Sec3d}
The three dimensional case offers a simpler laboratory
where to test these ideas before tackling the physically relevant 4d case.
In spin foams, there are two main reasons for the simplicity of the 3d case.

The first one is the solidity of the conjecture that the spin foam model implements the dynamics.
For 3d Riemannian quantum gravity, a body of evidence supports a specific model, known as
the Boulatov group field theory \cite{Boulatov}. When the generated 2-complex is dual to 
a Regge triangulation, the related kernel is the old Ponzano-Regge model \cite{Ponzano} 
(or the Turaev-Viro \cite{Turaev} for the case with cosmological constant). 
This solidity is two-fold.
On the one hand, the boundary Hilbert space of this model can be identified
with the canonical loop quantization of 3d GR \cite{Noui}. On the other hand, it can be seen
as a discrete version of Witten's quantization \cite{Witten}.\footnote{This
correspondence is particularly clear for non-zero $\Lambda$, since both Turaev-Viro
and Witten models are the square of an SU(2) Chern-Simons theory \cite{Witten, Roberts}.
For a recent nice argument that extends this correspondence to the $\Lambda=0$ case, see \cite{Barrett3d}.}

The second reason is that the spin foam
dynamical variables entering the sums have a direct metric meaning: on a fixed triangulation,
the spin foam amplitude is a sum over edge lengths, in terms of which the discrete metric can be
straighforwardly expressed (as it is done e.g. in Regge calculus).

\subsection{Ponzano-Regge model}

Let us for the moment ignore the summation over 2-complexes, and 
consider simply a given Regge triangulation. We also take a vanishing cosmological constant.
On a fixed triangulation the Ponzano-Regge kernel is a sum over edge lengths, with
amplitude given by a product of Wigner's 6j symbols 
associated to each tetrahedron,
\equ\label{K3d}
K[s] = \sum_{j_e} \prod_e (-1)^{2j_e} d_{j_e} \prod_\tau (-1)^{\sum_{e\in\tau}j_e} \{6j\}.
\nequ
In this expression $d_j=2j+1$ and the sum is over the internal spins only, with the external ones
fixed to the value assigned by the spin network $s$. The edge lengths are given in terms of the spins
by $\ell_e = \lp(j_e+\f12)$. The latter was taken as an ansatz in the original Ponzano-Regge paper,
but it is precisely what emerges from the canonical loop quantum gravity spectral analysis. 

Like \Ref{Kgb}, this expression needs gauge-fixing, as was realized already by Ponzano and
Regge (see also \cite{Ooguri}). The regularizing factor was later interpreted in \cite{FreidelD} as a division by the volume of diffeomorphisms of the triangulation, and a Faddeev-Popov procedure
introduced. This has been recently improved in \cite{Barrett3d}.
It fixes the same gauge freedom discussed above for linearized Regge calculus. 

The key for the semiclassical limit is the fact that Regge calculus
emerges in the large spin limit. This is due to the asymptotic behaviour of the 6j symbol,
given by the famous Ponzano-Regge formula \cite{Ponzano, Woodward}
\eqa\label{asymp}
(-1)^{\sum_{e}j_e} \{6j\} &=& \f1{\sqrt{12\, \pi\, V(\ell_e)}} \, 
{\cos\left( \f1{\lp} S_{\rm R}[\ell_e] +\f\pi 4  \right)} \no && +o({j^{-\f52}}),
\neqa
where $V(\ell_e)$ is the volume of the tetrahedron with edge lengths $\ell_e$ and $S_{\rm R}[\ell_e]$ the Regge action.
This fact is at the basis of the correct behaviour of $n$-point functions \cite{3d,3d1,3d2}, and can also be used
to study semiclassical properties of geometry and matter coupling \cite{Hackett,Speziale}.

From \Ref{asymp} it follows that the large spin limit of \Ref{K3d} is\footnote{Discarding the irrelevant 
$\f\pi4$ term. The extra edge phases in \Ref{K3d} provide the $2\pi$ factors in the
standard definition $\eps_e  = 2\pi -\sum_{\tau\in e} \theta_e^\tau(\ell_e)$ of the deficit angle.}  
\equ\label{K3dR}
K[q] = \sum_{\{\eps_\tau\} = \pm} \int \prod_e d \ell_e \, \mu(\ell_e) \, \prod_\tau  
e^{\f{i}{\lp} \eps_\tau S_{\rm R}(\ell_e)},
\nequ
where $q$ is the boundary geometry described in terms of the boundary edge lengths 
and the measure is (up to a numerical factor)
$\mu(\ell_e) = \prod_e \ell_e \prod_\tau V_\tau{}^{-1/2}$.
Similar measures have been considered also in quantum Regge calculus (see e.g. \cite{HamberM}),
where one starts directly from \Ref{K3dR} (without the $\eps_\tau$ sums).
For the reader interested in comparing the two approaches, I would like to stress
two important differences at this level. 
First of all, in Regge calculus one has to add by hand
constraints to impose the triangle inequalities, which in turn guarantee the positivity of
the metric. 
Here the triangle inequalities are elegantly implemented in the $\{6j\}$ itself,\footnote{In a quantum sense: 
the Clebsch-Gordan conditions satisfied by the $\{6j\}$ are somewhat looser than the real triangle
inequalities; in particular there are (measure zero) configurations of admissable spins with non-positive $V^2$.} 
which in turns fixes the power of the volume term in $\mu(\ell_e)$.
But more importantly, the measure described above is uniquely selected by the requirement 
of invariance under Pachner moves. 
As the measure is responsible for quantum corrections, a dynamical and unique choice
is crucial to make perturbation theory predictive.
The use of Pachner moves
to prove triangulation independence of the model and uniquely select the measure
is a novelty of the spin foam formalism, and brings a new light into this
long standing issue of quantum Regge calculus. 

Notice the $i$ in the exponents of  \Ref{K3dR}. Unlike in Regge calculus and conventional QFT, this is present
regardless of the signature of spacetime. That is, both in the SU$(2)$-based Euclidean Ponzano-Regge model \Ref{K3dR},
and in its Lorentzian version, based upon SU$(1,1)$ and discussed for instance in \cite{FreidelLor}.
Conventionally one defines these oscillating integrals through an analytic continuation of time to imaginary values.
As well known, a background independent framework as the one considered here does not allow us to do
so. An alternative prescription consists of adding a $+i \vareps$ term to the action,
similarly to the Feynman prescription in the proper time representation of the free scalar field propagator,
\equ
\f{i}{p^2 + i \vareps} = \int_0^\infty dN e^{i N (p^2 + i\vareps)}.
\nequ
This is the prescription that we take in the following, with the caveat that its physical viability 
is still to be demonstrated\footnote{It has been argued that the right Wick rotation for gravity comes from
analytically continuing the lapse function, $N\mapsto i N$. Freidel has pointed out that this can be 
achieved adding the term $+i\vareps \int \sqrt{g}$ to the action \cite{FreidelWick}.}.
This prescription has the advantage of being background-independent, but now
the Euclidean theory is not anymore directly related to the Lorentzian one. 
At least at this stage of understanding, this prescription and the Euclidean models should be mostly seen as toy models.

\subsection{Edge correlations}
The leading order in the $\lambda$ expansion of the Boulatov group field theory is a 2-complex
with a single 4-valent vertex. This trivial 2-complex is dual to a tetrahedron, and its boundary spin network can only have one graph. 
Consequently in the expression \Ref{Wsf} the summation over the graphs is dropped,
and one is left only with the summation over the spins (in a 3d Regge triangulation 
there are no SU$(2)$-intertwiner degrees of freedom),
\equ\label{W3d}
W^q_{abcd} = \sum_j {\mathbbm h}_{ab}(j) {\mathbbm h}_{cd}(j) \Psi_q(j) K[j] 
\nequ
where $K[j]= \prod_e d_{j_e} \, \{6j\}$.
This expression defines the leading order in $\lambda$ of the graviton propagator. 
In the 3d case we can, without loss of generality, look only at the components $W_{ab} \equiv W_{aabb}$.
In the canonical theory, ${\mathbbm h}_{aa}(j)$ is realized as (the fluctuations of) an edge length,
thus $W_{ab}$ gives edge correlations.

To fix ideas, take $q$ to be the metric on a 
2-sphere. Its simplest non-degenerate discretization 
has four equilateral triangles, each side of length say $\ell_0 = \lp (j_0+\f12)$.
We can refine the discretization of the boundary increasing the number of triangles;
however as the group field theory only generates 4-valent vertices, a finer triangulation of the boundary will
necessarily have more vertices inside, and thus be a higher order term in $\lambda$.
In this sense, the group field theory expansion controls the refinement of the boundary triangulation.

At fixed order of $\lambda$, we can study the perturbative expansion in $\lp$ (large spin expansion), around $q$.
The leading order contributes to the linearized theory.
Unfortunately, at this stage we do not have a prescription to compute the semiclassical state $\Psi_q[j]$ 
from the dynamics (see \cite{bs} for a discussion), not even for the linearized theory.
Hence we are not in a situation to make predictive calculations. The attitude is more to choose
a specific model and test whether it is viable at all, assuming an appropriate boundary state exists.
To proceed we make the ansatz that this state can be approximated at leading order by a Gaussian peaked
on a chosen classical geometry $q$ for the tetrahedron. 
Denote $\alpha$ the matrix in the Gaussian. Ideally, the entries of this matrix should be fixed 
if it is a solution of the linearized dynamics. Furthermore, we encode the $+i \vareps$
prescription here, by taking Re $\alpha >0$. The perturbative expansion is defined, at lowest order, using
\Ref{asymp} in \Ref{W3d}, and then expanding the Regge action around the background $q$.
With Re $\alpha >0$ the integrals are well defined, and the
leading order of the large spin expansion, corresponding to the free propagator (at this order
in $\lambda$), is simply the second momentum of a Gaussian integral. The kinetic 
matrix of this Gaussian integral is $\alpha + i G$, with $G$ the
Hessian of the Regge action on a single tetrahedron:
\equ\label{Wlo3d}
W_{ab} = \f{4}{j_0^2} \Big(\alpha + i G \Big)^{-1} + o(\lp) + o(\lambda),
\nequ
where the prefactor comes from the field insertions. 
This is the leading order contribution in $\lambda$ to the free graviton propagator.
The loop corrections $o(\lp)$ can be computed 
from the exact expression \Ref{W3d}. 
The calculations become 
increasingly intricate as the $\lp$ order is increased, but the set-up for
the full perturbative expansion is perfectly well-defined \cite{3d2}.
The $\lambda$ corrections can be computed from spin foams with more vertices,
see next Section.

Is the result \Ref{Wlo3d} in any way consistent with what we know from (Riemannian) linearized quantum gravity?
A direct comparison is obscured by its complex nature
in spite of the Riemannian signature. The Riemannian signature here is just a toy model,
and not the analytic continuation of the physical signature. A more meaningful comparison would be
between a Lorentzian spin foam model and the physical propagator. Nevertheless, there are many things
that can be learned from this toy model. 
First of all, recall that the propagator is a gauge dependent quantity, in particular in 3d is a pure gauge quantity. 
Where is the dependence on the gauge in \Ref{Wlo3d}?
As it turns out, $G$ on its own is not invertible, due to the diffeomorphism invariance of the Regge action  \cite{Dittrich}; thus the term $\alpha$ coming from the boundary state must act as a gauge-fixing term, in order for the inverse matrix \Ref{Wlo3d} to be well-defined.

Both $\alpha$ (see \cite{Rovelli,3d,semi}) 
and $G$ scale like $1/j_0$, so we obtain $W\sim1/j_0$. As in 3d loop gravity $\lp j_0$
is a distance, \Ref{Wlo3d} has the right scaling to be consistent with general relativity in
the low-energy limit. 
However a real consistency check requires more than the scaling: the free graviton propagator 
is a precise function of the distance between the two points, and it has a well-defined tensorial structure. 

How can this information be contained in the right hand side of \Ref{Wlo3d}? 
How is $\alpha$ encoding the gauge-fixing and the asymptotic flatness necessary to make sense of the standard graviton?
Even more stringently, how can it make sense to compute the graviton propagator on a single tetrahedron?
And how are the higher orders in $\lambda$ going to reproduce the continuum limit?

I claim that an appropriate choice of $\Psi_q[j]$ exists such that these questions can be 
answered in the positive. To understand how this happens,
it is useful to make a little detour and gain some insight from quantum Regge calculus. There one can show
that a $\Psi_q[\ell]$ exists such that the equivalent of \Ref{W3d} gives the right lattice propagator.
To see this, consider the single tetrahedron as part of an infinite lattice of flat 3d space. 
For each tetrahedron $\tau$, take the matrix $G_\tau$ of second derivative of the Regge action evaluated on the
flat background. By choosing a rectangular lattice (so that each edge of any tetrahedron is different)
we can keep a certain generality for $G_\tau$ without losing the simplicity of working with a flat background.
The action $S_{\rm L}\equiv \f1{16\pi G_{\scr N}}\sum_\tau G_\tau$ can be shown \cite{Dittrich} to reduce in the continuum limit to general relativity linearized around a flat background.
The Euclidean quantum theory is constructed as the path integral with weight 
$e^{-S_{\rm L}}$, and related to the Lorentzian theory by a Wick rotation in the background time.
The 2-point function gives the correct free graviton propagator on the lattice
(including the correct distance dependence, right tensorial structure, and correct pure-gauge nature) 
which reduces to the standard one in the continuum limit. 

To make contact with the spin foam calculation,
a formula analog to \Ref{W3d} (with integrals instead of sums) 
can be obtained also in Regge calculus, performing the integration over the edges of the infinite
lattice in two steps: first, an integration over the infinite lattice outside a chosen tetrahedron,
which defines $\Psi_q$ precisely as in \Ref{P}. Second, the remaining integration on the edges of the 
tetrahedron. The final result is the same lattice propagator, however we can stop at the first step and 
look at the characteristics of $\Psi_q$. 
In particular, its evaluation requires fixing the asymptotic boundary conditions and the gauge.
The result is indeed a Gaussian, with $\alpha$ uniquely fixed by the dynamics, the boundary conditions and 
the gauge choice. With this $\alpha$ (and without the $i$) \Ref{Wlo3d} gives the correct lattice free propagator.

This analysis performed in \cite{Dittrich} teaches us that the right propagator only emerges if the boundary state
is unique, and selected by the dynamics. Conversely,
it suggests that it makes sense to study the low-energy limit
of the theory using \Ref{W3d} and \Ref{Wsf}, and hints to some of the properties 
that the boundary state needs to have. At first order in $\lambda$, the (trivial, since we are in 3d) dynamics, gauge-fixing and $+i \vareps$ prescription are all encoded in the boundary state,
through a specific form of the matrix $\alpha$.

This said about the meaningfulness of \Ref{W3d} and the semiclassical limit, let me comment on the literature, where a version of this model has been studied extensively \cite{3d,3d1,3d2}.
Recall that the boundary state should carry the gauge-fixing. In \cite{3d,3d1}, part of the gauge is fixed by fixing four of the edge lengths, and only the correlation between two (opposite) edges is studied. 
Furthermore, the (now two by two) $\alpha$ matrix is taken to be diagonal.
This unphysical simplification does not change the qualitative picture, which is anyway the only aspect we can
study at the moment: quantitative statements will require a unique dynamical selection of the boundary state
and the inclusion of the sum over 2-complexes.
On the other hand, this simplification turned out to be very useful from the point of view of the numerical analysis,
and offered the possibility to investigate possible non-perturbative 
boundary states in terms of Bessel functions \cite{3d1}.

The numerical results support all the analytic calculations done so far, and were also used to investigate
a number of side issues. For instance a class of measures with different powers of $d_j$ was considered in \cite{3d1}, showing that the one required by triangulation independence minimizes the magnitude of the corrections.
Numerical work in progress includes the extension of \Ref{W3d} to more than one tetrahedron, thus addressing directly
the questions raised above:
Does the boundary state truly peak on a classical solution in the bulk?
Does it induce a unique orientation on it?
These issues might be clear at the perturbative level, but we need to have some control also at the
non-perturbative level. They are currently under investigation \cite{Costas}.
From this viewpoint notice also the phase factors in \Ref{K3d}, needed for the triangulation independence.
They are not present in the initial literature \cite{3d,3d1,3d2}
which uses a single tetrahedron, but are crucial to study more than one tetrahedron.

\subsection{Summing over the triangulations}\label{SecTr}
After discussing what could be the relation between the calculation on a fixed triangulation and the lattice
graviton, the next key question is how to recover the continuum limit when including higher orders in $\lambda$,
thus allowing arbitrarily fine triangulations.

Above I argued that the theory can reproduce the lattice graviton if a physical state
with specific semiclassical properties exists. To emphasize the fact that a
fixed triangulation $\Delta$ (dual to $\ga$) was used, I now redub that state $\Psi_{q_\Delta}[\ga, j]$.
In the previous Section, we used a Gaussian ansatz and the experience from lattice Regge calculus, 
we were able to establish some of the properties needed for the function
$\Psi_{q_\Delta}[\ga, j]$ peaked on the discrete classical geometry $q_{\Delta}$ (intrinsic and extrinsic). 
To include the sum over 2-complexes, we need to be able to write, within group field theory,
a function of a continuum metric $g$ peaked on the continuum geometry $q$. This should formally
resemble some sort of averaged state over all the graphs $\ga$ of the classical metric $q$
represented on $\ga$, schematically
\equ\label{PsiGen}
\Psi_q[g] = \mean{ \Psi_{q_{\ga}}[\ga,j]}_{\ga}.
\nequ 
This is of course a much harder problem than finding just $\Psi_{q_\Delta}$.
The existence of such a state is key to the study of the semiclassical limit and the extraction of
physical predictions from the theory.

Furthermore, there is another aspects of this state that will be crucial for future developments.
Notice in fact that the boundary state is the only quantity in \Ref{W3d} carrying a scale.
On a fixed graph such as the previous tetrahedal one, the scale was set by $j_0$. In the general 
expression \Ref{PsiGen}, the scale is set by two quantities: the number of vertices in the 
dominant graph $\ga_0$, and the average size of a cell $j_0$.
Thus $\Psi_q[g]$ carries a scale $\mu=(\ga_0, j_0)$, which depends upon the choice of $q$ and 
on the state being physical. This scale can be seen as the analog of the energy scale of the external legs
in conventional $n$-point functions in momentum space.
If we were able to write down such a state, then this framework could be accessible to the techniques of
renormalization theory. In particular, we could investigate the possibility of resumming the contribution
of subdominat graphs in a renormalization $\lambda(\mu)$ of the group field theory coupling constant.

The investigation of the role of classical solutions in group field theory is still at a preliminary
stage \cite{Fairbairn}, but I stress that this is a fundamental direction to explore.

\section{Four dimensional case}\label{Sec4d}
The most studied model was introduced by Barrett and Crane ten years ago \cite{bc}, and
gives the kernel as a sum over areas of the triangulations, similarly to what happens in
the Ponzano-Regge model. 
The kernel is given by another SU(2)-invariant object, briefly denoted 10j symbol for its functional
dependence. 
Its large spin limit gives a Regge-like action, 
where now the areas $A_t = \lp^2 (2j+1)$ are the independent variables \cite{Barrett4d}.
Schematically (for details see \cite{Barrett4d, Baez, Rovelli}), the behaviour on a single 4-simplex is
\equ\label{asymp4d}
\{10j\} \sim P(A_t) \, {\cos\left( \f1{\lp^2} \, S_{\rm R}[A_t] \right)}.
\nequ
The understanding of the one-loop measure $P$ is very limited, and further given the 
general limitations of the model, I will overlook a number of 
details.\footnote{For the experts, I would like to mention a point that I find somewhat amusing.
The measure term $P$ scales like $j^{-9/2}$. On the other hand,
in a 4-simplex there are ten areas, so the one-loop factor is (one over the square root of)
the determinant of a ten by ten matrix, which would lead to a scaling 
$j^{-10/2} \sim V^{-5/2}$ which is the scaling conjectured by Misner \cite{Misner} in '57. 
The missing factor of $j$ in $P$ comes, as in the 3d case, from the presence of a zero mode
in the Regge action for a 4-simplex.
More recent proposals for the measure include \cite{DeWitt} or \cite{Bern}.} 
This asymptotic behaviour is reminiscent of the key property supporting the validity of the Ponzano-Regge model in 3d.
Despite this and other good properties, this model (i) does not match loop gravity on its boundary states,
and (ii) does not have the right semiclassical limit \cite{Alesci}.
While the first shortcoming might not constitute per se evidence against this model, the second one
is sufficient to rule it out.
Intuitively, the difficulties with this model can be traced back to the fact that the areas of a triangulation
are not good metric variables, as opposed to the edge lengths. They only are 
on a single 4-simplex, and even there there are singular configurations where the same set of areas
corresponds to inequivalent discrete metrics \cite{BarrettA}. Far from being irrelevant,
this type of configurations occurs for instance in a regular hypercubical lattice, the most
natural setting to study perturbations around flat spacetime.
Hence the only setting where the Barrett-Crane model can capture general relativity is on a single
4-simplex around a non-singular configuration, for instance in the equilateral one.

This strongly limitates the interest in this model. Yet the model is non-trivial under many perspectives
and offers a good example to test the ideas described so far and, 
at least in the restricted equilateral setting, see qualitatively what
4d loop quantum gravity could dynamically be like.
This is the setting that has been investigated so far in the 
literature \cite{Rovelli, Bianchi, Livine, Chris, BianchiModesto}.

The $W_{ab}$ components correspond to area correlations, and the remaining to angle correlations.
Let us look at $W_{ab}$ first.
Consider $q$ the metric on a 3-sphere, discretized at leading order in $\lambda$ as
an equilateral 4-simplex with areas proportional $j_0$.
Restricting to the perturbative expansion around this equilateral configuration,
the situation is then similar to the 3d case:
inserting \Ref{asymp4d} in the 4d version of \Ref{W3d}
we can compute the leading order as a Gaussian integral:
it has precisely the same structure of \Ref{Wlo3d} but now
as a ten by ten matrix, representing the correlations among the ten areas on the 3d boundary of a 4-simplex.
As in 4d $j_0$ is a distance squared, we get a scaling consistent with linearized quantum gravity.

Again the $\lp$ perturbative expansion can be computed analytically, but more interesting
is to use the exact definition to investigate numerically the full quantum gravity effects at small scales/high energies.
This was possible thanks to the efforts of Christensen \cite{Chris} and adaptive 
Monte Carlo methods (VEGAS algorithm). Due to the complexity of the Barrett-Crane kernel,
we were only able to extract reliable simulations in the unphysical case of diagonal $\alpha$,
which however does not affect the qualitative picture.

The numerical results, appeared in \cite{Chris}, 
confirmed the analytic evaluation of the leading order, but also led to the
remarkable result that the typically divergent behaviour is dynamically regularized:
the analysis showed the presence of a peak close to the Planck
scale, as shown in the attached figure.
\begin{figure}[ht]
\includegraphics[width=8.3cm]{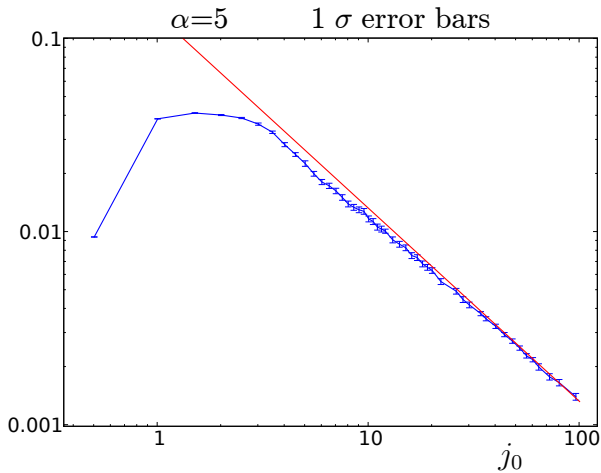}
\begin{picture}(0,0)(0,0)\put(-50,-2){\large$j_0$}\end{picture}
\caption{\small{Numerical study of \Ref{W} (dots), versus the analytic result
of the leading order, on a log-log plot.
Raw data and more plots are available at
\mbox{\texttt{http://jdc.math.uwo.ca/graviton}}}}\label{fig}
\end{figure}
The presence of this peak is a purely non-perturbative effect, and its exact location depends
on the value of $\alpha$.
We interpret the presence of a peak as follows. The discrete microscopic structure of
the theory provides a trivial regularization at any scale, in the sense that
only half-integer steps in the distance between the two points are allowed. 
This is a direct consequence of the kinematical discreteness of the spectra, and not a dynamical effect.
The non-trivial effect of the short scale dynamics is to introduce a suppression of the correlations,
which instead of increasing monotonically at shorter distances, reach a maximum and then decrease
(the effect being less noticeable for small values of $\alpha$).
In this sense, the divergent behaviour of the
graviton correlations gets regularized at high energies by the discrete structure of spin foams.
This shows how the full theory might enhance the effective field theory where the latter breaks down,
and it confirms the intuition that spacetime can not be considered as fluctuating around the flat metric 
at the Planck scale.

This picture can be taken seriously only if the semiclassical limit of the theory consistently
reproduces general relativity. The correct scaling is
a positive indication, but a satisfactory answer requires the correct inverse squared distance
behaviour, and the tensorial structure of a spin 2 massless particle.
The 3d case discussed above suggests to use the emergence of Regge calculus in \Ref{asymp4d} to support this result.
Indeed, 4d quantum Regge calculus reproduces linearized quantum gravity in the continuum limit \cite{Rocek}.
However, the fundamental variables of Regge calculus are the edge lengths, in terms of which we can
directly express the metric. Conversely, the fundamental variables of the Barrett-Crane model
are the areas of the triangles. 
As anticipated above, the areas are not good metric variables: a collection of 
4-simplex contributions like the Regge-like action of \Ref{asymp4d} on an infinite lattice does not
reproduce classical general relativity in the continuum limit. It has been
suggested that the right action can be obtained adding constraints for the areas \cite{Makela}, 
but these would be non-local on the 4-simplex, extremely hard to write explicitly, 
and further would not solve the issue of singular configurations.

Recently Dittrich and myself proposed an alternative solution: instead of constraining the areas, 
we added angles between triangles as variables to the action. 
In \cite{Dittrich2} we showed that these extra variables solve naturally the problem of defining a 
discrete metric uniquely, when satisfying constraints which are local and easy to write explicitly.
Using these constraints we wrote an action which is completely equivalent to Regge's, 
and thus to general relativity in the continuum limit.
If our action emerges from a spin foam model, then the same argument that we run in the previous Section for the
Ponzano-Regge model can be used, and support the idea that with the correct boundary state
the model has the right low-energy physics, at least on a fixed triangulation.

Due to the extra angle variables, our action can arise 
only from spin foam models which sum over both spins and intertwiners, not only spins like
in Barrett-Crane.
In turn, precisely the lack of intertwiner degrees of freedom was taken by Rovelli and his
group as the fundamental problem with the Barrett-Crane model. This shows up explicitly when trying
to compute the angle correlations, which fail to have even just the correct $1/j_0$ scaling \cite{Alesci}.
Following this line of thought, an improved spin foam model was proposed in \cite{EPR}, which
at least naively has the right matching with spin network states on the boundary. Not long afterwards,
Livine and myself realized \cite{noi2} that this model can also be obtained starting from the coherent states 
we had previously introduced in \cite{noi1}, a result independently found also by Krasnov and Freidel \cite{loro}.
Further developments include \cite{Sergei, EPR2, EPRL, altri}, and excitement has arisen around the possibility
that this new model might indeed cure the problems of Barrett-Crane's and give general relativity in the 
semiclassical limit. 
The failure of the Barrett-Crane model teaches us that the emergence of 
a Regge-like equation from the non-perturbative kernel is not enough to guarantee the
correctness of the classical equations of motion. Unless the dynamical variables are the edge
lengths, additional terms to the action have to be expected. In their absence, the model
is likely not to have the right semiclassical limit.

For the new model, an explicit asymptotic formula like \Ref{asymp4d} is not known, thus we cannot 
study the semiclassical limit through the perturbative expansion of correlators as described above.
Promising preliminary results have appeared, both analytically and numerically \cite{Magliaro1, Conrady08, Khavkine, ABMP}. In particular Conrady and Freidel showed \cite{Conrady08} that the amplitude of the vertex is
dominated by configurations of the variables which have a metric interpretation. Althought they were not
able to compute an explicit asymptotic formula like \Ref{asymp4d}, their results are very encouraging.

Among the remarkable improvements to the Barrett-Crane model, the 
new model (in both original and coherent state versions) can also accomodate 
the Immirzi parameter $\ga$ \cite{loro, EPRL}, which is crucial to make contact with loop gravity.
A non-trivial feature of this extension is that the two versions remain equivalent for $\ga<1$,
whereas they differ for $\ga>1$. Somewhat surprisingly, the case $\ga=1$ can not be covered by either version.
This case is special because the action reduces to the Plebanski action in terms of self-dual 
two-forms \cite{Pleb}.
For this special case only, an alternative to the Barrett-Crane model had been already proposed by Reisenberger \cite{Mike1}.
The spin foam representation of this model is less simple and elegant than Barrett-Crane's, a feature
that makes it less amenable to a group field theory description and to the calculation of
correlators as described in this review.
However the semiclassical limit of the kernel can be studied in the group representation,
and on a single 4-simplex a simplicial version of general relativity remarkably emerges \cite{Mike2}. 
This allows us for instance to see explicitly how the extrinsic curvature is 
incorporated in SU(2). See \cite{Mike1, Mike2} for details.
Thanks to these properties I believe it is certainly worth investigating more
this alternative model.

\section{Discussion}\label{SecConcl}

I described how the conventional definition of correlators in quantum field theory can be  
manipulated to accomodate a general background-independent formalism.
This manipulation does not address per se the open problems of the continuum theory, such as
the ambiguities in the measure or the UV divergences. It offers, on the other hand, a possibility
to study the correlators in a background-independent theory such as loop quantum gravity,
and test whether those difficulties can be overcome.

The immediate relevance of the technique for loop quantum gravity is that it offers a way to compute
perturbatively $n$-point functions. This is a key step towards the 
extraction of physical predictions from the theory.
In particular, application of this technique to the Barrett-Crane spin foam model in 3+1 dimensions
has shown that the model fails to reproduce general relativity in
the semiclassical limit. This might look at first sight like a disappointing result, in particular spoiling the 
nice picture of short scale dynamical regularization of the correlations described above.
On the contrary, I think it is a remarkable success:
the application of the general boundary correlations to study
the semiclassical limit has led us to rule out the most studied spin foam model!
Such an outcome is encouraging for the robustness of this way to extract physics
and for the falsifiability of loop gravity in general.
Furthermore, this negative result has triggered 
new efforts towards the understanding of the spin foam dynamics,
and for the first time in ten years we have new proposals to improve it.\footnote{I 
am optimistic that the qualitative picture of dynamical
regularization in Barrett-Crane found in \cite{Chris} might survive in the correct model: results from a number
of directions suggest that its origin might not be so much in the facets of the model,
but rather in the fact that the fundamental variables of the quantum theory are holonomies, 
instead of the classical connection. This feature is shared by all these models.}
The new models \cite{EPR,noi1,noi2,loro} might lead to a vertex amplitude with a better-behaved low-energy dynamics.

Once a valid vertex amplitude is secured, I believe three key steps have to be taken before this way 
of extracting physics can be made predictive.

1. \emph{The continuum limit}. On a fixed triangulation, the correctness of the semiclassical limit
of spin foams can be related to the emergence of Regge calculus. To study the continuum limit, we
need to allow the triangulation to be arbitrarily fine. This in turn requires including all the terms in
the perturbative expansion in $\lambda$ of the group field theory. However, as we discussed here, the technique
to study $n$-point functions naturally introduces a scale fixed by the semiclassical geometry of the
observer. This scale is characterized by two quantities, the refinement of the boundary triangulation,
and the average size of its cells. It is reasonable to expect that such a scale determines which interpolating
spin foams are dominant in the summation. It is less obvious, but would be extremely important, that the contributions from subdominant spin foams can be taken into account through a renormalization of $\lambda$.

2. \emph{The semiclassical boundary state}. The boundary state is the key ingredient that makes this
technique work, both on a fixed triangulation and, probably even more profoundly through renormalization,
in the continuum limit. In the explicit calculations appeared in the literature, we assume that the
boundary state is peaked on a classical intrinsic and extrinsic geometry, and we use it to fix the
gauge of the graviton propagator and to encode the Wick rotation. If the theory turns out not to possess
states with these properties, the whole technique might be jeopardized. Pursuing this line of
research is even more crucial as the explicit knowledge of the states is mandatory to make the technique truly predictive.

3. \emph{The extension to Lorentzian models}. This is a smaller step than the previous two.
In principle, we already know how to deal with Lorentzian signature, to which both the Barrett-Crane model
and the new models can be extended. The principal difference with the Euclidean 
signature is the non-compact nature of the gauge group. A lesson to learn from this line of investigation
is the viability or not of the Wick rotation proposed here, and eventually the suggestion of alternatives.

\subsubsection*{Acknowledgements}
I would like to thank Dario Benedetti, Bianca Dittrich, Razvan Gurau and Matteo Smerlak for valuable discussions.
This research was supported by Perimeter Institute for Theoretical Physics.  
Research at Perimeter Institute is supported by the Government of Canada through 
Industry Canada and by the Province of Ontario through the Ministry of Research \& Innovation.

\newpage


\end{document}